\documentclass[nofootinbib,prd,aps,twocolumn,showpacs,tightenlines,floatfix]{revtex4}

\usepackage{graphicx}
\newcommand{\beq}{\begin{equation}}
\newcommand{\eeq}{\end{equation}}
\newcommand{\bqa}{\begin{eqnarray}}
\newcommand{\eqa}{\end{eqnarray}}
\newcommand{\cj}[1]{{\bf [j]}}
\newcommand{\cm}[1]{{\bf [m]}}

\def\square{\vcenter{\vbox{\hrule height.4pt
          \hbox{\vrule width.4pt height8pt
          \kern8pt\vrule width.4pt}\hrule height.4pt}}}

\def\sumint{\hbox{$\sum$}\!\!\!\!\!\!\int}

\begin{document}

NORDITA-2004-47, TUW-04-15\\

\title{Three-loop $\Phi$-derivable Approximation in QED}
\author{Jens O. Andersen}
\affiliation{Nordita, Blegdamsvej 17, DK-2100 Copenhagen, 
Denmark\footnote{e-mail: jensoa@nordita.dk}}

\author{Michael Strickland}
\affiliation{Institut f\"ur Theoretische Physik, Technische Universit\"at Wien,
        Wiedner Hauptstrasse 8-10, A-1040 Vienna, 
Austria~\footnote{e-mail: mike@hep.itp.tuwien.ac.at}
\\ (\today)
}

\begin{abstract}
{\footnotesize 
In this paper we examine $\Phi$-derivable approximations in QED.
General theorems tell us that the gauge dependence of the 
$n$-loop $\Phi$-derivable approximation shows up at order $g^{2n}$
where $g$ is the coupling constant.
We consider the gauge dependence of the 
two-loop $\Phi$-derivable approximation to the Debye mass 
and show that it is of order $e^4$ as expected.
We solve the three-loop $\Phi$-derivable approximation in QED
by expanding sum-integrals in powers of $e^2$ and $m/T$, where 
$m$ is the Debye mass which satisfies a variational gap equation.
The results for the pressure and the Debye mass are accurate to order $e^5$.

}
\end{abstract}
\maketitle

\section{Introduction}
\begin{figure}[htb]
\includegraphics[width=7.7cm]{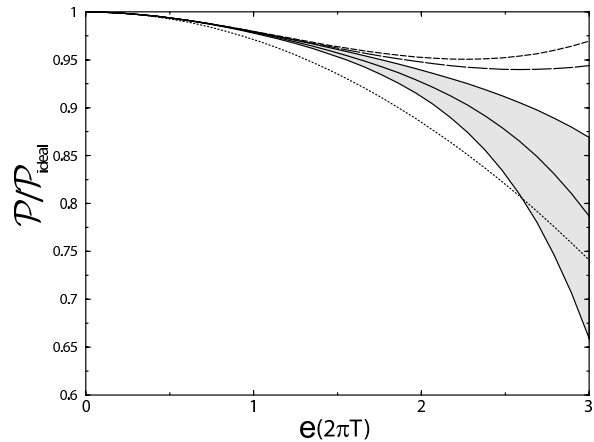}
\caption{Weak-coupling expansion for the pressure to orders $e^2$ 
(dotted curve), $e^3$ (dashed curve),
$e^4$ (long-dashed curve), and $e^5$ (solid lines+band)
normalized to that of an ideal gas as a function of
$e(2\pi T)$.}
\label{pert}
\end{figure}

The thermodynamic functions for hot field theories can be calculated as
a power series in the coupling constant $g$ at weak coupling.
The free energy has been calculated through
to order $g^4$ in~\cite{wrong,arnold1} for scalar $\phi^4$ theory,
in~\cite{qed4} for QED and in~\cite{arnold1} 
for nonabelian gauge theories.
The corresponding calculations to order $g^5$ were carried out 
in Refs.~\cite{singh,ea1}, Refs.~\cite{parwani,Andersen} and
Refs.~\cite{zhai,ea2}, 
respectively.
In Fig.~\ref{pert}, we show the successive perturbative approximations to
${\cal P}/{\cal P}_{\rm ideal}$ as a function of $e(2\pi T)$. Each partial
sum is shown as a band obtained by varying the renormalization scale $\mu$ 
by a factor
of two around the central value $\mu=2\pi T$.
We have only done this variation for the $e^5$ approximation.
To express $e(\mu)$ in terms of
$e(2\pi T)$, we use the 
solution to the one-loop renormalization group equation in QED.
The Figure shows that the weak-coupling expansion is poorly convergent unless
the coupling constant is small and that it is very sensitive to the
renormalization scale $\mu$. The lack of convergence seems to be related
with screening and quasiparticles which is associated with the soft
momentum scale of order $eT$. The instability of the weak-coupling
expansion is a generic problem in hot field theories and makes it essentially
useless for quantitative predictions.

There are several ways of systematically 
reorganizing the perturbative expansion to improve its convergence 
properties and various approaches have been discussed in detail
in the review papers Refs.~\cite{birdie,kraemmer, review}. 
One of these methods is {\it screened perturbation theory} (SPT)
which in the context of thermal field theory was 
introduced by Karsch, Patk\'os and Petreczky.~\cite{K-P-P-97}
(See also refs.~\cite{yuk,steve,kleinert}).
In this approach, one introduces a single variational parameter
which has a simple interpretation as a thermal mass.
In SPT a mass term is added to and subtracted from the scalar Lagrangian
with the added piece kept as part of the free Lagrangian and the
subtracted piece associated  with the interactions.
The mass parameter satisfies a variational equation which is obtained
by the principle of minimal sensitivity.

In gauge theories, one cannot simply add and subtract a local mass
term as this would violate gauge invariance.
Instead one adds and subtracts to the Lagrangian a hard thermal loop
(HTL) improvement term.
The free part of the Lagrangian then includes the HTL self-energies and 
the remaining terms are treated as perturbations.
{\it Hard thermal loop perturbation theory} 
is a manifestly gauge invariant approach that can
be applied to static as well as dynamic quantities.
SPT and HTL perturbation theory have been applied to three and 
two loops~\cite{spt,AS-01,htl1,fermions,htl2,aps1}, 
respectively, and the convergence properties are improved
dramatically compared to the weak-coupling expansion.

The $\Phi$-derivable approach is another way of 
reorganizing the perturbative expansion which is variational of nature.
In this approach, one uses
the exact propagator as a variational function.
Its formulation
was first constructed by Luttinger and Ward~\cite{lw} and by Baym~\cite{baym}.
Later it was generalized to relativistic field theories by Cornwall, Jackiw
and Tomboulis~\cite{CJT-74}. The approach is based on the fact that the
thermodynamic potential can be expressed
in terms of the two-particle irreducible (2PI)
effective action which has a diagrammatic expansion
involving the 2PI skeleton graphs.  Although here we focus
on equilbrium physics we note that the 2PI formalism and its generalizations are 
also very useful when studying non-equilibrium real-time physics
\cite{Aarts:2002dj,Aarts:2003bk,Berges:2004pu}.

The $\Phi$-derivable approach has 
several attractive features. One is that it
respects the global symmetries of the
theory. Thus it is consistent with
the conservation laws that follow from the Noether's theorem.
Second, when evaluated at the stationary point, one is guaranteed 
thermodynamic consistency~\cite{baym}.
Finally, it turns out that 
the two-loop $\Phi$-derivable approximation has an additional property.
The entropy reduces to the one-loop expression at the variational point.
This property was first shown for QED by Vanderheyden and Baym~\cite{van} and
later generalized to QCD by Blaizot, Iancu and 
Rebhan~\cite{bir1,bir2,rebb}.

Applying the $\Phi$-derivable approach to quantum field theories, one
is facing two nontrivial issues. The first issue is
the question of renormalization. The three-loop calculations by Braaten
and Petitgirard~\cite{ep1} for a massless scalar field theory indicate
that there are ultraviolet divergences at order $g^6$ that cannot
be eliminated by any renormalization of the
coupling constant. These calculations seem to 
contradict the results from the papers by van Hees 
and Knoll~\cite{knoll}, 
and by Blaizot, Iancu and Reinosa~\cite{urko}, which show that the
2PI effective action can be systematically renormalized.

The second issue is that of gauge-fixing dependence. While the exact
2PI effective action is gauge independent at the stationary point, 
this property is often lost in approximations.
The problem has recently been examined by 
Arrizabalaga and Smit~\cite{arri}, who showed that the $n$-loop 
$\Phi$-derivable approximation $\Phi_n$,
which is defined by the truncation of the
action functional after $n$ loops, has a gauge dependence that shows up 
at order $g^{2n}$.
Furthermore, if the $n$th order solution to the gap
equation is used to evaluate the complete effective action, the 
gauge dependence first shows up at order $g^{4n}$.
For a general proof of the gauge invariance of the exact 2PI effective
action we refer the reader to Ref.~\cite{Carrington:2003ut}.

In gauge theories, approximate solutions to the 
gap equations in
two-loop $\Phi$-derivable
approximation in terms of HTL self-energies have been obtained by
Blaizot, Iancu and Rebhan~\cite{bir1,bir2,rebb}, and 
by Peshier~\cite{peshier}.
For hard external momentum, these solutions are obtained by evaluating the
one-loop self-energy diagrams with bare propagators.
For soft external momentum, the solution is simply the HTL self-energies, 
which is 
in the imaginary-time formalism reduces to the Debye mass.
The resulting approximation reproduces the pressure to order $g^3$.

Finally, we mention that a dimensionally reduced version of the 
three-loop $\Phi$-derivable approximation was solved in the case of scalar
field theory~\cite{review}.
This was done by treating the contribution to the pressure from the
nonzero Matsubara modes in strict perturbation theory 
and applying the 
$\Phi$-derivable approach to an effective three-dimensional field theory
for the zero-frequency mode
that has been obtained by dimensional reduction.
The results are comparable to those obtained
in SPT~\cite{spt} and the $\Phi$-derivable approximation in 3+1 
dimensions~\cite{ep1}.

In this paper, we solve the three-loop $\Phi$-derivable approximation
for QED. This is done by applying the strategy developed 
in Ref.~\cite{ep1}. It consists of expanding the
sum-integrals systematically in powers of $e$ and $m/T$
where $m$ is a variational mass parameter of order $eT$.
Our result for the pressure reproduces the weak-coupling expansion
to order $g^5$ and is gauge invariant.

The paper is organized as follows.
In Sec.~II, we briefly discuss the application of the $\Phi$-derivable
approach to QED and the general framework developed in Ref.~\cite{ep1}
to solve it systematically. In Sec.~III, we solve the two-loop 
$\Phi$-derivable approximation and discuss the issue of gauge dependence.
In Sec.~IV, we solve the three-loop $\Phi$-derivable approximation.
We summarize and draw some conclusions in Sec.~V. 
There are two appendices where our 
notation and conventions are given and 
where we list the sum-integrals and integrals that are needed.

\section{$\Phi$-derivable approximations}
In  this section, we briefly discuss the 2PI effective action formalism
and $\Phi$-derivable approximations.

The Euclidean Lagrangian of massless QED is
\bqa
{\cal L}&=&{1\over4}F_{\mu\nu}^2+\bar{\psi}\gamma_{\mu}D_{\mu}\psi+
{\cal L}_{\rm gf}\;,
\eqa
where $F_{\mu\nu}=\partial_{\mu}A_{\nu}-\partial_{\nu}A_{\mu}$
is the field strength tensor,
$D_{\mu}=\partial_{\mu}+ieA_{\mu}$ is the covariant derivative, and $e$
is the electric coupling.
${\cal L}_{\rm gf}$ is the gauge-fixing part of the Lagrangian.
In general covariant gauge, 
the gauge-fixing part of the Lagrangian is
\bqa
{\cal L}_{\rm gf}&=&{1\over2\xi}\left(\partial_{\mu}A_{\mu}\right)^2\;.
\eqa 
In the remainder of this Sec. and in Sec.~III, we keep $\xi$
general in order to discuss the problem of gauge dependence. In Sec.~IV,
we specialize to Feynman gauge ($\xi=1$), which by far is the easiest
gauge for practical calculations.

\begin{figure}[htb]
\includegraphics[width=5.2cm]{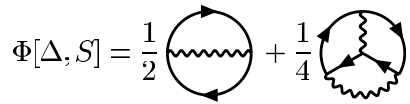}
\caption{$\Phi$-derivable two- and three-loop skeleton graphs.}
\label{graphs}
\end{figure}

\begin{figure}[htb]
\includegraphics[width=6.2cm]{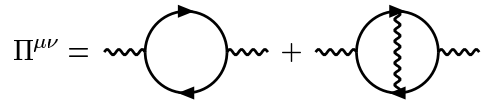}
\caption{One- and two-loop photon self-energy graphs.}
\label{selfgraphs1}
\end{figure}

\begin{figure}[htb]
\includegraphics[width=7.2cm]{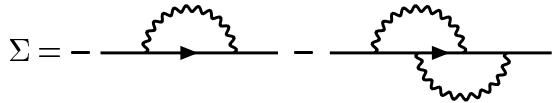}
\caption{One- and two-loop electron self-energy graphs.}
\label{selfgraphs2}
\end{figure}

The thermodynamic potential $\Omega$ of QED is
\bqa\nonumber
\Omega[\Delta,S]&=&{1\over2}{\rm Tr}\log\Delta^{-1}-{\rm Tr}\log S^{-1}
-{\rm Tr}\log\Delta^{-1}_{\rm gh}
\\ &&\nonumber
-{1\over2}{\rm Tr}\,\Pi\Delta
+{\rm Tr}\,\Sigma S
+{\rm Tr}\,\Pi_{\rm gh}\Delta_{\rm gh}
\\ &&
+\Phi[\Delta,S]\;,
\label{phidef}
\eqa
where 
$\Delta_{\mu\nu}(P)$ and $S(P)$ is the exact photon and electron propagator,
respectively, and $\Delta_{\rm gh}(P)$ is the propagator for the ghost.
$\Pi_{\mu\nu}(P)$ is the polarization tensor and $\Sigma(P)$ 
is the electron self-energy. We can then write
\bqa
\Delta_{\mu\nu}^{-1}(P)&=&
\left[\Delta^0_{\mu\nu}(P)\right]^{-1}
+\Pi_{\mu\nu}(P)\;,\\
S^{-1}(P)&=&/\!\!\!\!P+\Sigma(P)\;,
\eqa 
where $\Delta^0_{\mu\nu}(P)$ is the free propagator in covariant gauge:
\bqa
\Delta_{\mu\nu}^0(P)={\delta_{\mu\nu}\over P^2}
-(1-\xi){P_{\mu}P_{\nu}\over P^4}\;.
\eqa
The trace in Eq.~(\ref{phidef}) is over Dirac and Lorentz indices
as well as space-time.
In covariant gauges, the ghost field decouples from the other fields and 
so the ghost self-energy $\Pi_{\rm gh}(P)$ vanishes 
identically~\footnote{In nonabelian gauge theories, the ghost does not
decouple in covariant gauges which makes the calculation significantly
more involved. In Ref.~\cite{rebb} the authors are employing 
the temporal axial gauge in which the ghost does decouple. However, there
are other problems with this gauge at finite 
temperature~\cite{landshoff,bellac}.}.
The functional $\Phi[\Delta,S]$ is the sum of all two-particle irreducible vacuum diagrams.
We define the $n$-loop $\Phi$-derivable approximation
$\Omega_n$ to the thermodynamic potential $\Omega$
as the truncation of the action functional
after $n$ loops.
The two-particle irreducible vacuum diagrams are shown diagrammatically in 
Fig.~\ref{graphs} up to three-loop order. 
The corresponding self-energies that are obtained by cutting a line, are shown
in Figs.~\ref{selfgraphs1} and~\ref{selfgraphs2}.

The exact propagators satisfy the variational
equations
\bqa
\label{var1}
{\delta\Omega[\Delta,S]\over\delta \Delta}&=&0\;,\\
\label{var2}
{\delta\Omega[\Delta,S]\over\delta S}&=&0\;.
\eqa
Using Eq.~(\ref{phidef}), the variational equations~(\ref{var1}) 
and~(\ref{var2}) can be written as
\bqa
\label{var3}
\Pi_{\mu\nu}(P)&=&2{\delta\Phi[\Delta,S]\over\delta\Delta_{\mu\nu}(P)} \;,\\
\Sigma(P)&=&-{\delta\Phi[\Delta,S]\over\delta S(P)} \;.
\label{var4}
\eqa

In QED, we know that thermal fluctuations generate a mass $m$ for the zeroth
component of the gauge field $A_0$
which is of order $eT$ and screens the interactions.
The strategy for solving the $n$-loop $\Phi$-derivable approximation
is to introduce a mass variable which is of order $eT$ and then calculate
the sum-integrals as double expansions in $e^2$ and $m/T$.
This strategy was developed in Ref.~\cite{ep1} 
in order to
solve the three-loop $\Phi$-derivable approximation in scalar field theory.
It turns out that the gap equations~(\ref{var3}) and~(\ref{var4})
have a recursive structure that allows us to solve for their 
dependence of the external momentum $P$. We follow Braaten and Petitgirard
and choose the Debye mass as the mass parameter.
The Debye mass is the solution to the equation
\bqa
p^2+\Pi_{00}(0,{\bf p})&=&0\;,\hspace{1cm}p^2=-m^2\;.
\label{debyedef}
\eqa
In the variational equations~(\ref{var3}) and~(\ref{var4}),
there are two important mass scales. One is {\it soft} and is 
of order $eT$. This scale is set by the Debye mass $m$. 
The other is the {\it hard} scale of order $2\pi T$ and is set by the
nonzero Matsubara modes.
We will assume the coupling is sufficently small so that the scales
$m$ and $2\pi T$ are well separated. This allows one to expand the 
sum-integrals in powers of $e^2$ and $m/T$. The gap equations will then
be solved in the two momentum regions separately.
For hard momentum $P$, we expand the 
polarization tensor as follows:
\bqa\nonumber
\Pi_{\mu\nu}(P)&=&e^2
\Pi_{\mu\nu}^{2,0}(P)
+e^4\left[\Pi_{\mu\nu}^{4,0}(P)
\right.\\ &&\left.
+\Pi_{\mu\nu}^{4,1}(P)+...
\right]+...\;,
\label{exp1}
\eqa
where $\Pi_{\mu\nu}^{n,k}(P)$ is of order $T^2(m/T)^{k}$. Similarly, the
electron propagator is expanded as
\bqa\nonumber
\Sigma(P)&=&e^2\left[
\Sigma^{2,0}(P)+\Sigma^{2,1}(P)
+...
\right]
\\ &&
+e^4\left[
\Sigma^{4,0}(P)+\Sigma^{4,1}(P)
+...
\right]+...\;,
\label{exp2}
\eqa
where $\Sigma^{n,k}(P)$ is of order $T^2(m/T)^k$. 
For soft momentum $P=(0,{\bf p)}$, we expand the 
longitudinal part of polarization tensor as 
follows~\footnote{Since the infrared limit of the other components
of $\Pi_{\mu\nu}(0,{\bf p})$ 
vanishes, the corresponding contribution 
to the free energy also vanishes.}:
\bqa\nonumber
\Pi_{00}(0,{\bf p})&=&m^2+e^2\left[
\sigma^{2,0}(p)+\sigma^{2,2}(p)+...
\right]
\\ &&
+e^4\left[
\sigma^{4,0}(p)+...
\right]+...\;,
\label{exp3}
\eqa
where $\sigma^{n,k}(p)$ is of order $m^2(m/T)^{k}$.

For hard momentum, we can expand ${1\over2}{\rm Tr}\log\;\Delta^{-1}$ about the
free propagator, since the self-energy is perturbative corrections starting
at order $e^2$. This yields
\bqa\nonumber
{1\over2}{\rm Tr}\log\Delta^{-1}&=&
{1\over2}(d+1)\sumint_P\log P^2
+{1\over2}e^2\sumint_P
{\Pi_{\mu\mu}^{2,0}(P)\over P^2}
\\ && 
\hspace{-2.0cm}
+{1\over2}e^4\sumint_P\left[
{\Pi_{\mu\mu}^{4,0}(P)\over P^2}
-{1\over2}{\Pi_{\mu\nu}^{2,0}(P)\Pi_{\mu\nu}^{2,0}(P)\over P^4}
\right]+...
\;.
\label{hard}
\eqa
The gauge-dependent terms in Eq.~(\ref{hard}) drop out since
the photon self-energy in QED is transverse to all orders:
\bqa
P_{\mu}\Pi_{\mu\nu}(P)&=&0\;.
\eqa

For soft momentum, the expansion is
\bqa\nonumber
{1\over2}{\rm Tr}\log\Delta^{-1}&=&
{1\over2}T\int_p\log\left(p^2+m^2+e^2\sigma_{2,0}(p)+...\right)\\ 
&&
\hspace{-2.3cm}
=\;
{1\over2}T\int_p\log\left(p^2+m^2\right)
+{1\over2}e^2T\int_p{\sigma_{2,0}(p)\over p^2+m^2}
+...\;,
\label{softie}
\eqa
Again we have used the transversality of the photon propagator to eliminate
the gauge-dependent terms. We do not need the expansion of ${\rm Tr}\,\Pi D$
since we will use the gap equation to eliminate this term.

We also need the expansions for ${\rm Tr}\log S^{-1}$ and 
${\rm Tr}\,\Sigma S$.
Since the electron momentum is always hard, we can expand about the
free propagator and obtain
\bqa\nonumber
{\rm Tr}\log S^{-1}&=&2\sumint_{\{P\}}\log P^2
\\ && \nonumber
+e^2\sumint_{\{P\}}{\rm Tr}\left[
{\Sigma^{2,0}(P)/\!\!\!\!P\over P^2}
+{\Sigma^{2,1}(P)/\!\!\!\!P\over P^2} +...\right]
\\ 
&& 
\hspace{-0.9cm}
-{1\over2}e^4\sumint_{\{P\}}\left[
{\Sigma^{2,0}(P)/\!\!\!\!P\Sigma^{2,0}(P)/\!\!\!\!P\over P^4}+...\right]+...
\;,
\label{h1}
\\ \nonumber
{\rm Tr}\,\Sigma S&=&e^2\sumint_{\{P\}}{\rm Tr}
\left[{\Sigma^{2,0}(P)/\!\!\!\!P\over P^2}
+{\Sigma^{2,1}(P)/\!\!\!\!P\over P^2}+...\right]
\\ &&
\hspace{-0.7cm}
-e^4\sumint_{\{P\}}\left[{\Sigma^{2,0}(P)/\!\!\!\!P\Sigma^{2,0}(P)/\!\!\!\!P
\over P^4}+...\right] +...
\;,
\label{h2}
\eqa
where the trace on the right-hand side is only over Dirac indices.

The contribution from the ghost field is as usual
\bqa
{\rm Tr}\log\Delta_{\rm gh}^{-1}=\sumint_P\log P^2\;.
\eqa

By inserting the expansions for the self-energies into the gap equations
and expanding systematically in powers of $e$ and $m/T$, we obtain
expressions for $\Pi_{\mu\nu}^{n,k}(P)$, $\Sigma^{n,k}(P)$
and $\sigma_{\mu\nu}^{n,k}(p)$.
By matching coefficients of $e^n$ on both sides and solving the equations
simultaneously and recursively. 

\section{Two loops}
In the two-loop $\Phi$-derivable approximation, there is only a single
diagram contributing to $\Phi[D,S]$ which is the left diagram in 
Fig.~\ref{graphs}.
The two-loop thermodynamic potential $\Omega_2$ is
\bqa\nonumber
\Omega_2[\Delta,S]&=&{1\over2}{\rm Tr}\log\Delta^{-1}-{\rm Tr}\log S^{-1}
-{\rm Tr}\log\Delta^{-1}_{\rm gh}
\\ && \nonumber
-{1\over2}{\rm Tr}\,\Pi\Delta
+{\rm Tr}\,\Sigma S
+{1\over2}e^2\sumint_{P\{Q\}}
\\ &&
\times{\rm Tr}
\left[S(Q)\gamma^{\mu}S(P+Q)\gamma^{\nu}\Delta_{\mu\nu}(P)\right]\;.
\label{2lps}
\eqa
 The gap equations are obtained by varying the thermodynamic potential
$\Omega_2$ with respect to $\Pi_{\mu\nu}(P)$ and $\Sigma(P)$:
\bqa
\label{pigap1}
\Pi_{\mu\nu}(P)&=&e^2\sumint_{\{Q\}}
{\rm Tr}\left[S(Q)\gamma^{\mu}S(P+Q)\gamma^{\nu}\right]\;,\\
\Sigma(P)&=&e^2\sumint_{Q}\gamma^{\mu}S(P+Q)\gamma^{\nu}\Delta_{\mu\nu}(Q)\;.
\label{pigap2}
\eqa
It follows from the coupled gap equations that both 
$\Pi_{\mu\nu}(P)$ and $\Sigma(P)$ 
are nontrivial functions of the external momentum $P$.

The gap equation~(\ref{pigap1}) can be used to simplify 
equation~(\ref{2lps}) for
$\Omega_2$:
\bqa\nonumber
\Omega_2&=&{1\over2}{\rm Tr}\log\Delta^{-1}
-{\rm Tr}\log\Delta^{-1}_{\rm gh}
-{\rm Tr}\log S^{-1} 
\\ &&
+{\rm Tr}\,\Sigma S\;.
\label{2luss}
\eqa
Substituting the expansions for the various terms into~(\ref{2luss})
and truncating at the appropriate order, we obtain 
\bqa\nonumber
\Omega_2&=&{1\over2}(d-1)\sumint_P\log P^2+{1\over2}T\int_p\log(p^2+m^2)
\\ &&
-2\sumint_{\{P\}}\log P^2
+{1\over2}e^2\sumint_P{\Pi_{\mu\mu}^{2,0}(P)\over P^2}\;.
\eqa
We note that the function $\Sigma^{2,0}(P)$ drops out.
In the two-loop $\Phi$-derivable approximation, we only need the trace
of $\Pi_{\mu\nu}^{2,0}(P)$, while in the three-loop
$\Phi$-derivable approximation, we need the function itself.

The solution to the gap equations for hard momentum to order $ e^2$ are 
obtained by
using bare propagators in the loops.
In this manner, we find
\bqa\nonumber
\Pi_{\mu\nu}^{2,0}(P)&=&\sumint_{\{Q\}}\left[
{8Q_{\mu}Q_{\nu}\over Q^2(P+Q)^2}-{4\delta_{\mu\nu}\over Q^2}
\right.\\ &&\left.
\hspace{-1cm}
+{2P^2\delta_{\mu\nu}\over
Q^2(P+Q)^2}+{4P_{\mu}Q_{\nu}+4P_{\nu}Q_{\mu}\over Q^2(P+Q)^2}
\right]\;.
\eqa

We also need to solve the gap equation for soft momentum in order to
determine the Debye mass.
Through order $e^3$, the longitudinal part of 
polarization tensor at zero frequency reads
\bqa
\Pi_{00}(0,{\bf p})&=&m^2\;,
\eqa
where the Debye mass is
\bqa\nonumber
m^2&=&-4(d-1)e^2\sumint_{\{Q\}}{1\over Q^2}\,\\
&=&{16\pi^2\over3}\alpha T^2\;,
\eqa
where $\alpha=e^2/(4\pi)^2$.
The thermodynamic potential through order $e^3$ then reduces 
to
\bqa\nonumber
\Omega_2&=&{1\over2}(d-1)\sumint_P\log P^2
+{1\over2}T\int_p\log(p^2+m^2)
\\ && \nonumber
-2\sumint_{\{P\}}\log P^2
-(d-1)e^2
\\ &&
\times
\left[\sumint_{P\{Q\}}{2\over P^2Q^2}-\sumint_{\{PQ\}}{1\over P^2Q^2}
\right]\;.
\eqa
Using the expressions for the integrals and sum-integrals in the appendices,
this reduces to
\bqa
\Omega_2&=&-{11\pi^2T^4\over180}\left[1-{50\over11}\alpha
+{320\sqrt{3}\over33}\alpha^{3/2}
\right]\;.
\label{2loop}
\eqa
Eq.~(\ref{2loop}) agrees with the weak-coupling result through order 
$e^3$~\cite{parwani}. Thus the two-loop $\Phi$-derivable approximation
sums up the leading contribution from the plasmon diagrams.

We close this section by discussing the problem of gauge dependence that arises
when going beyond order $e^3$. For example, to 
calculate the Debye mass
to order $e^4$, we need 
to include the function
$\Sigma^{2,0}(P)$ in the dressed electron propagator on the right-hand side 
of the gap equation~(\ref{pigap1}). 
This function is
\bqa
\nonumber
\Sigma^{2,0}(P)&=&(1-d)
\sumint_Q{/\!\!\!\!\!\!\!\left(P+/\!\!\!\!Q\right)\over Q^2(P+Q)^2}
-(1-\xi)
\\ &&
\hspace{-1.2cm}
\label{hardel}
\times\sumint_{Q}
\left[{/\!\!\!\!Q\over Q^4}-{/\!\!\!\!P\over Q^2(P+Q)^2}
-{P^2/\!\!\!\!Q\over Q^4(P+Q)^2}\right]
\;.
\label{softel}
\eqa
The function $\Sigma^{2,0}(P)$ arises from hard photon momenta in the
self-energy graph in Fig.~\ref{selfgraphs2}.
We note that it is gauge dependent and this is
due the photon line in the one-loop self-energy graph
shown in Fig.~\ref{selfgraphs2}.
Since $\Sigma^{2,0}(P)$ is gauge dependent, this introduces
a gauge dependence at order $e^4$ in the longitudinal
part of polarization tensor. At
zero frequency, one finds:
\bqa\nonumber
\Pi_{00}(0,{\bf p})&=&
-4(d-1)e^2\sumint_{\{Q\}}{1\over Q^2}
\\ && \nonumber
+{2\over3}(d-1)e^2p^2
\sumint_{\{Q\}}{1\over Q^4}
\\ &&\nonumber\hspace{-1.2cm}
+{8\over3}(d-1)(d-3)e^4
\sumint_{\{Q\}}{1\over Q^4}\left[
\sumint_{R}{1\over R^2}-\sumint_{\{R\}}
{1\over R^2}
\right]
\\ && \nonumber\hspace{-1cm}
+8\left(1-\xi\right)e^4\sumint_{\{P\}Q}
\left[
{P_0^2\over P^2Q^4(P+Q)^2}-{2P_0^2\over P^4Q^4}
\right]\;.
\\ 
\label{fake}
\eqa
The equation for $\Pi_{00}(0,{\bf p})$ is ultraviolet divergent and requires
renormalization. The divergence proportional  to $p^2$ is removed by
wave-function renormalization in the usual manner (see also Sec.~\ref{3lps}).
The other divergence which arises from the last line in Eq.~(\ref{fake})
can only be removed by a gauge-dependent renormalization
of the coupling constant~\footnote{Only in Feynman gauge is the
polarization tensor finite after wave-function renormalization.}. 
This would lead to a gauge-dependent 
gap equation and Debye mass. 
Eq.~(\ref{fake}) depends on the gauge-fixing parameter, but this 
is not in contradiction with the fact that the photon
self-energy is manifestly gauge invariant. The point is that in the two-loop
$\Phi$-derivable approximation, we are not including all contributions
to $\Pi_{\mu\nu}(P)$ of order $e^4$. This is done when one considers the
three-loop approximation and we have explicitly checked that the gauge
dependence cancels algebraically as we include the two-loop self-energy
graph in Fig.~\ref{selfgraphs1}.

\section{Three loops}
\label{3lps}

The three-loop $\Phi$-derivable approximation to the free energy is
\bqa\nonumber
\Omega_3[\Delta,S]&=&
{1\over2}{\rm Tr}\log\Delta^{-1}-{\rm Tr}\log S^{-1}
-{\rm Tr}\log\Delta^{-1}_{\rm gh}
\\ && \nonumber \hspace{-3mm}
-{1\over2}{\rm Tr}\,\Pi\Delta
+{\rm Tr}\,\Sigma S
\\ && \nonumber \hspace{-3mm}
+{1\over2}e^2\sumint_{P\{Q\}}{\rm Tr}
\left[S(Q)\gamma^{\mu}S(P+Q)\gamma^{\nu}\Delta_{\mu\nu}(P)\right]
\\ && \nonumber \hspace{-3mm}
+{1\over4}e^4\sumint_{P\{QR\}}{\rm Tr}
\left[S(Q)\gamma^{\mu}S(R)\gamma^{\alpha}
S(R-P)
\right.\\ && \nonumber
\left.
\times
\gamma^{\nu}S(Q-P)\gamma^{\beta}\Delta_{\mu\nu}(P)\Delta_{\alpha\beta}(Q-R)
\right]\;.
\\ &&
\label{3phi}
\eqa
The gap equations are again obtained by varying~(\ref{3phi})
with respect to the photon and electron self-energies:
\bqa\nonumber
\Pi_{\mu\nu}(P)&=&
e^2\sumint_{\{Q\}}
{\rm Tr}\left[S(Q)\gamma^{\mu}S(P+Q)\gamma^{\nu}\right]\;,\\ \nonumber
&&+e^4\sumint_{\{QR\}}{\rm Tr}\left[
S(Q)\gamma^{\mu}S(R)\gamma^{\alpha}
\right.\\ &&\left.\nonumber
\times S(R-P)\gamma^{\nu}S(Q-P)
\gamma^{\beta}
\right]
\Delta_{\alpha\beta}(Q-R)\;,
\\ &&
\\ \nonumber
\Sigma(P)&=&e^2\sumint_{Q}\gamma^{\mu}S(P+Q)\gamma^{\nu}\Delta_{\mu\nu}(Q)
\\ && \nonumber
+e^4\sumint_{Q\{R\}}\gamma^{\mu}S(R)\gamma^{\alpha}S(R-Q)\gamma^{\nu}
\\ &&
\times S(P-Q)\gamma^{\beta}\Delta_{\alpha\beta}(Q)\Delta_{\mu\nu}(R-P)\;.
\label{3leself}
\eqa
The gap equation for $\Pi_{\mu\nu}(P)$ can be used to simplify the expression
for the thermodynamic potential:
\bqa\nonumber
\Omega_3[\Delta,S]&=&
{1\over2}{\rm Tr}\log\Delta^{-1}-{\rm Tr}\log S^{-1}
-{\rm Tr}\log\Delta^{-1}_{\rm gh}
\\ && \nonumber
+{\rm Tr}\Sigma S
\\ &&\nonumber
-{1\over4}e^4\sumint_{P\{QR\}}{\rm Tr}\left[S(Q)\gamma^{\mu}S(R)\gamma^{\alpha}
S(R-P)
\right.\\ && \left.
\times
\gamma^{\nu}S(Q-P)\gamma^{\beta}\Delta_{\mu\nu}(P)\Delta_{\alpha\beta}(Q-R)
\right] .
\label{3phinew}
\eqa
Substituting the expansions~(\ref{hard})--(\ref{h2}) into~(\ref{3phinew}), 
we obtain
\bqa\nonumber
\Omega_3&=&
{1\over2}(d-1)\sumint_P\log P^2+{1\over2}T\int_p\log(p^2+m^2)
\\ && \nonumber \hspace{-6mm}
+{1\over2}e^2T\int_p{\sigma^{2,0}(p)\over p^2+m^2}
-2\sumint_{\{P\}}\log P^2
\\ && \nonumber \hspace{-6mm}
+{1\over2}e^2\sumint_P{\Pi_{\mu\mu}^{2,0}(P)\over P^2}
\\ && \nonumber \hspace{-6mm}
+{1\over2}e^4\sumint_P\left[{\Pi_{\mu\mu}^{4,0}(P)\over P^2}
+{\Pi_{\mu\mu}^{4,1}(P)\over P^2}
\right.\\ && \nonumber \left.
-{1\over2}\sumint_P{\Pi_{\mu\nu}^{2,0}(P)\Pi_{\mu\nu}^{2,0}(P)\over P^4}
\right]
\\ && \nonumber \hspace{-6mm}
-{1\over2}e^4\sumint_{\{P\}}
{\rm Tr}\left[
{\Sigma^{2,0}(P)/\!\!\!\!P\Sigma^{2,0}(P)/\!\!\!\!P
\over P^4}
\right.\\ \nonumber&&\left. 
+2{\Sigma^{2,0}(P)/\!\!\!\!P\Sigma^{2,1}(P)/\!\!\!\!P
\over P^4}
+{\Sigma^{2,1}(P)/\!\!\!\!P\Sigma^{2,1}(P)/\!\!\!\!P
\over P^4}
\right]
\\ && \nonumber \hspace{-6mm}
+{1\over2}(d-1)(5-d)e^4\sumint_{\{PQR\}}{1\over P^2Q^2R^2(P+Q+R)^2}
\\ &&\nonumber \hspace{-6mm}
+(d-1)(d-3)e^4\sumint_{PQ\{R\}}{1\over P^2Q^2R^2(P+Q+R)^2}
\\ && \nonumber \hspace{-6mm}
+8(d-1)e^4
\sumint_{\{Q\}R}{Q_0R_0\over Q^4R^2(Q+R)^2}
\left(T\int_p{1\over p^2+m^2}\right)
\;.
\\ &&
\label{0f}
\eqa
In the three-loop $\Phi$-derivable approximation, we only need the
trace of the functions $\Pi_{\mu\nu}^{4,0}(P)$ and $\Pi_{\mu\nu}^{4,1}(P)$.
These traces are significantly simpler to calculate than the 
functions themselves.
Furthermore, it turns out that the order-$e^4$ term in 
equation~(\ref{3leself}) for the
electron self-energy first contributes to the pressure at order $e^7$ and so
it is not needed.
In addition to the functions listed in Sec.~III, we need the 
following functions for hard $P$
\bqa
\Sigma^{2,1}(P)&=&
{2p_0\gamma_0-/\!\!\!\!P\over P^2}T\int_{q}{1\over q^2+m^2}\;, 
\label{1f}
\\ \nonumber
\Pi_{\mu\mu}^{4,0}(P)&=&
-4(d-1)^2\sumint_{\{Q\}R} \Biggl[
{1 \over Q^4 R^2} 
- {P^2 \over Q^4 R^2 (Q+R)^2}
\\ && \nonumber \hspace{-1.5cm}
+ {1 \over R^2 (P+Q)^2 (Q+R)^2}
- {2 P\cdot R \over Q^2 R^2 (P+Q)^2 (Q+R)^2} 
\Biggr]
\\ && \nonumber \hspace{-1.5cm}
-4(d-1)^2\sumint_{\{QR\}} \Biggl[
{P^2 \over Q^4 R^2 (P+Q)^2}
- {1 \over Q^4 R^2} 
\Biggr]
\\ && \nonumber \hspace{-1.5cm}
 -4 (d-1) \sumint_{\{QR\}}
\Biggl[
{ 4(Q\cdot R)^2 \over Q^2 R^2 (R-P)^2 (Q-P)^2 (Q-R)^2 }
\\ && \nonumber 
+ { P^4 \over Q^2 R^2 (R-P)^2 (Q-P)^2 (Q-R)^2 }
\\ && \nonumber 
- { 4 P^2 \over Q^2 R^2 (Q-P)^2 (Q-R)^2 }
\\ && \nonumber 
- {d-7\over2} { P^2 \over Q^2 R^2 (R-P)^2 (Q-P)^2 }
\\ && \nonumber 
+ (d-3) { 1 \over  R^2 (Q-P)^2 (Q-R)^2 }
\\ && \nonumber 
- { 2 Q^2 \over R^2 (R-P)^2 (Q-P)^2 (Q-R)^2 }
\\ && 
+ { 2 \over R^2 (R-P)^2 (Q-R)^2 }
\Biggr] \, ,
\\ \nonumber
\Pi_{\mu\mu}^{4,1}(P)&=&2(d-1)
\sumint_{\{Q\}}\left[
{2P^2\over Q^4(P+Q)^2}-{2\over Q^4}
\right.\\ &&\nonumber\left.
+{8q_0^2\over Q^6}
-{8P^2q_0^2\over Q^6(P+Q)^2}-{4p_0q_0P^2\over Q^4(P+Q)^4}
\right.\\ &&\left.
-{4P^2q_0^2\over Q^4(P+Q)^4}
\right]
\left(T\int_r{1\over r^2+m^2}\right)\;.
\label{3f}
\eqa
The functions $\Sigma^{2,1}(P)$ and $\Pi_{\mu\nu}^{4,1}(P)$ arise 
when the photon momentum in the relevant Feynman diagram is soft, while
$\Pi_{\mu\nu}^{4,0}(P)$ is when the photon momentum is hard.

Through order $e^5$, the longitudinal part of polarization tensor 
at zero frequency can be written as
\bqa\nonumber
\Pi_{00}(0,{\bf p})&=&
-4(d-1)e^2\sumint_{\{Q\}}{1\over Q^2}
\\ && \nonumber
+{2\over3}(d-1)e^2p^2\sumint_{\{Q\}}{1\over Q^4}
\\ &&\nonumber\hspace{-2cm}
+4(d-1)(d-3)e^4
\sumint_{\{Q\}}{1\over Q^4}\left[
\sumint_{R}{1\over R^2}-\sumint_{\{R\}}
{1\over R^2}
\right]
\\ &&
\hspace{-2cm}
+4(d-1)(d-3)e^4\sumint_{{\{P}\}}{1\over P^4}\;
\left(T\int_r{1\over r^2+m^2}\right)
\;.
\label{mbare}
\eqa

We have explicitly checked that~(\ref{mbare}) 
is independent of the gauge-fixing parameter $\xi$
in contrast to the gauge-dependent expression~(\ref{fake}).
Eq~(\ref{mbare}) can be rewritten as 
\bqa
\Pi_{00}(0,{\bf p})&=&m^2+e^2\sigma^{2,0}(p)\;,
\eqa
where the Debye mass satisfies
\bqa\nonumber
m^2&=&
-4(d-1)e^2\sumint_{\{Q\}}{1\over Q^2}
\\ && \nonumber
+{2\over3}(d-1)e^2p^2\sumint_{\{Q\}}{1\over Q^4}\bigg|_{p^2=-m^2}
\\ &&\nonumber\hspace{-2cm}
+4(d-1)(d-3)e^4
\sumint_{\{Q\}}{1\over Q^4}\left[
\sumint_{R}{1\over R^2}-\sumint_{\{R\}}
{1\over R^2}
\right]
\\ &&
\hspace{-2cm}
+4(d-1)(d-3)e^4\sumint_{{\{P}\}}{1\over P^4}\;
\left(T\int_r{1\over r^2+m^2}\right)
\;,
\label{mbare2}
\eqa
and
\bqa
\sigma^{2,0}(p)&=&{2\over3}(d-1)e^2(p^2+m^2)\sumint_{\{Q\}}{1\over Q^4}\;.
\label{s20}
\eqa
The gap equation~(\ref{mbare2}) is ultraviolet divergent and requires 
renormalization. 
The divergence is proportional to $p^2$
and is removed by wave-function renormalization:
\bqa
Z_A&=&1-{e^2\over12\pi^2\epsilon}\;.
\eqa
After having renormalized the static
polarization tensor $\Pi_{00}(0,{\bf p})$, 
the gap equation~(\ref{debyedef}) reduces to
\bqa\nonumber
m^2&=&{16\pi^2\over3}T^2\alpha\Bigg\{
1-{8\over3}\bigg[\log{\mu\over4\pi T}
+\gamma+2\log2+{7\over4}
\\ &&
-18\left({m\over4\pi T}\right)\bigg]\alpha
\Bigg\}\;.
\label{debyefinal}
\eqa
The result for the Debye mass~(\ref{debyefinal}) agrees with the
weak-coupling result~\cite{bip,Andersen} through order $e^5$.

Inserting the expressions~(\ref{1f})--(\ref{3f}) and~(\ref{s20})
into~(\ref{0f}),
the three-loop $\Phi$-derivable approximation then becomes
\bqa\nonumber
\Omega_3&=&
{1\over2}(d-1)\sumint_P\log P^2
+{1\over2}T\int_p\log(p^2+m^2)
\\ && \nonumber
-2\sumint_{\{P\}}\log P^2
-e^2(d-1)
\\ && \nonumber\times
\left[\sumint_{P\{Q\}}{2\over P^2Q^2}-\sumint_{\{PQ\}}{1\over P^2Q^2}
\right]\\ && \nonumber
+(d-1)^2e^4\sumint_{\{P\}}{1\over P^4}\left[
\sumint_{\{Q\}}{1\over Q^2}-\sumint_{Q}{1\over Q^2}
\right]^2\\ && \nonumber
-4(d-3)e^4\sumint_{P\{QR\}}{1\over P^4Q^2R^2}
\\ && \nonumber
-2(d-1)e^4\sumint_{PQ\{R\}}{1\over P^2Q^2R^2(P+Q+R)^2}
\\ && \nonumber
+{1\over2}(d^2-8d+11)e^4
\\ && \nonumber\times
\sumint_{\{PQR\}}{1\over P^2Q^2R^2(P+Q+R)^2}
\\ && \nonumber
-2(d-1)^2e^4\sumint_{\{P\}QR}{QR\over P^2Q^2R^2(P+Q)^2(P+R)^2}
\\ && \nonumber
-16e^4
\sumint_{P\{QR\}}{(QR)^2\over P^4Q^2R^2(P+Q)^2(P+R)^2}
\\ && \nonumber
-(d-1)(d-3)e^4\sumint_{\{P\}}{1\over P^4}
\left(T\int_q{1\over q^2+m^2}\right)^2\;.
\\ &&
\eqa
Note that we have kept a term that is proportional to $e^4m^2$
and first contributes at order $e^6$. This term arises from three-loop diagrams
where both photons are soft~\footnote{The complete $e^4m^2$ contribution
can also be obtained in a two-loop calculation using the dimensionally
reduced theory of QED (electrostatic QED) derived in~\cite{landsman,Andersen} 
.}. 
This contribution is manifestly gauge invariant
and we include this selective resummation in our final result.
It is interesting to note that the only contribution at order $e^5$
comes from the Debye mass; all the other contributions cancel algebraically

The expression for $\Omega_3$ is ultraviolet divergent. The divergences
can be eliminated by renormalizing the coupling constant.
This is done by the substitution $e^2\rightarrow Z^2_ee^2$, where
\bqa
Z_e^2&=&1+{e^2\over12\pi^2\epsilon}\;.
\label{from}
\eqa
Using the expressions for the sum-integrals and integrals listed
in the appendices,
we obtain
\bqa\nonumber
\Omega_3&=&-
{11\pi^2T^4\over180}\left\{1-{50\over11}\alpha
+{960\over11}\left({m\over4\pi T}\right)^3
\right.\\ &&\left.\nonumber
-\left[
{400\over33}\left(\log{\mu\over4\pi T}
+{3\over5}\gamma
-{2\over5}
{\zeta^{\prime}(-3)\over\zeta(-3)}
\nonumber
\right.\right.\right.\\ &&\left.\left.\left. \nonumber
+{4\over5}{\zeta^{\prime}(-1)\over\zeta(-1)}
+{319\over80}-{156\over25}\log2\right)
\right.\right.\\ &&\left.\left.
+{11520\over11}\left({m\over4\pi T}\right)^2
\right]\alpha^2
\right\}\;.
\eqa
Using the expression for the Debye mass in Eq.~(\ref{debyefinal}), one can
show that the three-loop $\Phi$-derivable approximation agrees with the
weak-coupling expansion through order $e^5$~\cite{parwani,Andersen}.

The renormalization group equation that follows from~(\ref{from}) is
\bqa
\mu{de^2\over d\mu}={e^4\over6\pi^2}\;,
\eqa
which coincides with the standard one-loop running of the coupling.

\begin{figure}[htb]
\includegraphics[width=7.7cm]{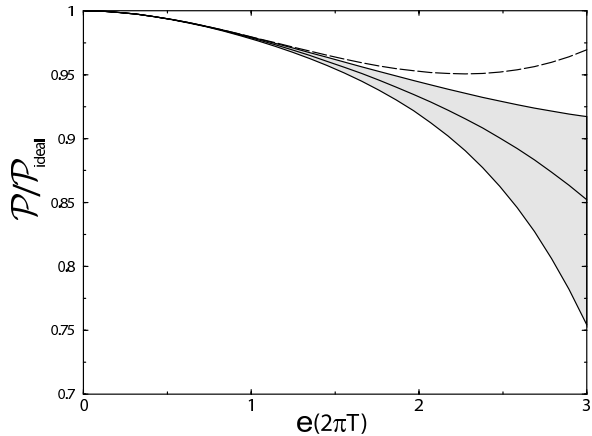}
\caption{Two- and three-loop $\Phi$-derivable approximations to the pressure
normalized to that of an ideal gas as a function of $e(2\pi T)$ shown as
dashed and solid lines, respectively. The three-loop
band is obtained by varying the renormalization scale, $\mu$, by a factor
of two around the central value $\mu=2\pi T$.  Note that the scale is
different than in Fig.~\ref{pert}.}
\label{qedfree}
\end{figure}

In Fig.~\ref{qedfree}, we show the two and three-loop 
$\Phi$-derivable approximations  
to the pressure normalized to that of an
ideal gas shown as dashed and solid lines, respectively. In the three-loop approximation, the band is obtained in the
usual manner by varying the renormalization scale $\mu$.
The three-loop band is slightly narrower when compared to the $e^5$-band
in Fig.~\ref{pert}. The approximations also seem to be slightly more 
stable than the successive weak-coupling approximations.  However, the
final result does not seem to be a dramatic improvement over the $e^5$
result.

\section{Summary}
In this paper we have solved the three-loop $\Phi$-derivable approximation
for the Debye mass and the free energy 
in QED by systematically expanding the sum-integrals in powers of
$e^2$ and $m/T$. 
The results are accurate to order $e^5$.
In the two-loop $\Phi$-derivable approximation, both 
the thermodynamic potential and gap equation are finite, and so
there is no running of the coupling. 
The solution to the gap equation is trivial; the Debye mass is given
by its weak-coupling expression.
This was also the case in the
two-loop calculation of Blaizot, Iancu and Rebhan in the case of 
QCD~\cite{bir1,bir2,rebb}.
In scalar field theory, the coupling is running incorrectly by a factor of
three~\cite{ep1,bir1,bir2,rebb}.
In the three-loop $\Phi$-derivable approximation, 
the expressions for the Debye mass and the 
free energy require wave-function renormalization and  renormalization of the
coupling constant, respectively. The running of the resulting 
renormalized coupling
agrees with the standard one-loop running in QED. 

We have also considered the problem of gauge dependence within the 2PI 
effective action formalism. We gave an 
explicit example of the how gauge dependence arises
in the one-loop gap equation for the photon propagator 
when one truncates the gap equations at order $e^4$.
Our calculation is in agreement with general results on the gauge dependence
of $\Phi$-derivable approximations~\cite{arri}. 

The method could also be used to solve the three-loop $\Phi$-derivable
approximation in QCD with an accuracy of order $g^5$; however, given
that the final three-loop $\Phi$-derivable result in QED does not
seem to dramatically improve the scale variation at large coupling
it is questionable whether this would be a worthwhile endeavor.
Going beyond three-loops in scalar theory as well as gauge theories
would be very difficult. One problem is that
there are new four-loop sum-integrals of order $g^6$ that have not yet
been evaluated. In nonabelian gauge theories, there is the additional
problem that the free energy at this order is sensitive to the
nonperturbative momentum scale $g^2T$ which is associated with 
screening of static magnetic fields.
This may cause the expansion of the three-loop $\Phi$-derivable
approximation to break down beyond order $g^5$.

\section*{Acknowledgments}
The authors would like to thank E. Braaten and A.K. Rebhan for 
useful discussions. 
M.S. was supported by FWF Der Wissenschaftsfonds Project M790-N08.

\appendix
\section{Sum-integrals}
In the imaginary-time formalism for thermal field theory, 
the 4-momentum $P=(P_0,{\bf p})$is Euclidean with $P^2=P_0^2+{\bf p}^2$. 
The Euclidean energy $p_0$ has discrete values:
$P_0=2n\pi T$ for bosons and $P_0=(2n+1)\pi T$ for fermions,
where $n$ is an integer. 
Loop diagrams involve sums over $P_0$ and integrals over ${\bf p}$. 
With dimensional regularization, the integral is generalized
to $d = 3-2 \epsilon$ spatial dimensions.
We define the dimensionally regularized sum-integral by
\bqa
  \hbox{$\sum$}\!\!\!\!\!\!\int_{P}& \;\equiv\; &
  \left(\frac{e^\gamma\mu^2}{4\pi}\right)^\epsilon\;
  T\!\!\!\!\!\!\sum_{P_0=2n\pi T}\:\int {d^{3-2\epsilon}p \over (2 \pi)^{3-2\epsilon}}\;,\\ 
  \hbox{$\sum$}\!\!\!\!\!\!\int_{\{P\}}& \;\equiv\; &
  \left(\frac{e^\gamma\mu^2}{4\pi}\right)^\epsilon\;
  T\!\!\!\!\!\!\sum_{P_0=(2n+1)\pi T}\:\int {d^{3-2\epsilon}p \over (2 \pi)^{3-2\epsilon}}\;,
\label{sumint-def}
\eqa

where $3-2\epsilon$ is the dimension of space and $\mu$ is an arbitrary
momentum scale. 
The factor $(e^\gamma/4\pi)^\epsilon$
is introduced so that, after minimal subtraction 
of the poles in $\epsilon$
due to ultraviolet divergences, $\mu$ coincides 
with the renormalization
scale of the $\overline{\rm MS}$ renormalization scheme.

\subsection{One-loop sum-integrals}
The specific one-loop sum-integrals needed are
\bqa
\sumint_{P}\log P^2&=&-{\pi^2T^4\over45}
\;,\\ \nonumber
\sumint_{P}{1\over P^2}
&=&{T^2\over12}
\left({\mu\over4\pi T}\right)^{2\epsilon}
\Bigg[1+\left(
2+2{\zeta^{\prime}(-1)\over\zeta(-1)}
\right)\epsilon\Bigg]\;,
\\ &&
\hspace{3.4cm}
\\  
\sumint_P {1 \over (P^2)^2} &=&
{1 \over (4\pi)^2} \left({\mu\over4\pi T}\right)^{2\epsilon} 
\,\left[ {1 \over \epsilon} + 2 \gamma
%
\right] \;, \\
\sumint_{\{P\}}\log P^2&=&{7\pi^2T^4\over360}
\;,\\ \nonumber
\sumint_{\{P\}}{1\over P^2}
&=&-{T^2\over24}
\left({\mu\over4\pi T}\right)^{2\epsilon}
\bigg[1
\\&& 
+\bigg(
2-2\log2
+2{\zeta^{\prime}(-1)\over\zeta(-1)}
\bigg)\epsilon
\bigg]\;,
\label{simple1}
\\ \nonumber
\sumint_{\{P\}}{1\over(P^2)^2}&=&
{1\over(4\pi)^2}\left({\mu\over4\pi T}\right)^{2\epsilon}
\bigg[
{1\over\epsilon}+2\gamma
+4\log2
\bigg]\;.
\\ &&
\eqa

The errors are all one order higher in $\epsilon$ than the smallest term 
shown. 
The calculations of these sum-integral is standard.
\subsection{Two-loop sum-integrals}
The two-loop sum-integrals that are needed all vanish:
\bqa
\sumint_{\{PQ\}}{1\over P^2Q^2(P+Q)^2}&=&0 \; , \\
\sumint_{PQ}{1\over P^2Q^2(P+Q)^2}&=&0 \; , 
\eqa

The errors are all of order $\epsilon$.
Details of the calculation of these two-loop sum-integrals can be found in
e.g. Ref.~\cite{arnold1}.

\subsection{Three-loop sum-integrals}
The three-loop diagrams needed are
\bqa\nonumber
\sumint_{PQR}{1\over P^2Q^2R^2(P+Q+R)^2}&=&
\\ && \nonumber
\hspace{-4.8cm}
{1\over(4\pi)^2}\left({T^2\over12}\right)^2
\left({\mu\over4\pi T}\right)^{6\epsilon}
\left[{6\over\epsilon}+{182\over5}
\right.\\ &&\left.
\hspace{-4cm}
-12{\zeta^{\prime}(-3)\over\zeta(-3)}
+48{\zeta^{\prime}(-1)\over\zeta(-1)}
\right]\;,\\ \nonumber 
\sumint_{\{PQR\}}{1\over P^2Q^2R^2(P+Q+R)^2}&=&
\\ \nonumber
\hspace{-3cm}
{1\over(4\pi)^2}\left({T^2\over12}\right)^2
\left({\mu\over4\pi T}\right)^{6\epsilon}
\left[{3\over2\epsilon}+{173\over20}
\right.\\ &&\left.
\hspace{-4.9cm}
-{63\over5}\log2
-3{\zeta^{\prime}(-3)\over\zeta(-3)}
+12{\zeta^{\prime}(-1)\over\zeta(-1)}
\right]\;,\\ 
\nonumber
\sumint_{PQ\{R\}}{1\over P^2Q^2R^2(P+Q+R)^2}&=&
\\&&\nonumber
\hspace{-4.8cm}
{1\over(4\pi)^2}\left({T^2\over12}\right)^2
\left({\mu\over4\pi T}\right)^{6\epsilon}
\left[-{3\over4\epsilon}-{179\over40}
\right.\\ &&\left.
\hspace{-4.8cm}
+{51\over10}\log2
+{3\over2}{\zeta^{\prime}(-3)\over\zeta(-3)}
-6{\zeta^{\prime}(-1)\over\zeta(-1)}
\right]\;,\\ \nonumber
\sumint_{\{P\}QR}{QR\over P^2Q^2R^2(P+Q)^2(P+R)^2}&=&
\\ &&\nonumber
\hspace{-4.8cm}
{1\over(4\pi)^2}\left({T^2\over12}\right)^2
\left({\mu\over4\pi T}\right)^{6\epsilon}
\left[{3\over8\epsilon}+{9\over4}\gamma+{361\over160}
\right.\\ && \left.
\hspace{-4.8cm}
+{57\over10}\log2
+{3\over2}{\zeta^{\prime}(-3)\over\zeta(-3)}
-{3\over2}{\zeta^{\prime}(-1)\over\zeta(-1)}
\right]\;, \\
\nonumber
\sumint_{P\{QR\}}{(QR)^2\over P^2Q^2R^2(P+Q)^2(P+R)^2}&=&
\\ && \nonumber
\hspace{-4.8cm}
{1\over(4\pi)^2}\left({T^2\over12}\right)^2
\left({\mu\over4\pi T}\right)^{6\epsilon}
\left[{5\over24\epsilon}+{1\over4}\gamma
\right.\\ && \left.\hspace{-5.4cm}
+{23\over24}
-{8\over5}\log2
-{1\over6}{\zeta^{\prime}(-3)\over\zeta(-3)}
+{7\over6}{\zeta^{\prime}(-1)\over\zeta(-1)}
\right]\;.
\eqa

The errors are all of order $\epsilon$.
The calculation of these three-loop sum-integrals was done in 
Ref.~\cite{arnold1} and details can be found there.

\section{Integrals}

Dimensional regularization can be used to
regularize both the ultraviolet divergences and infrared divergences
in 3-dimensional integrals over momenta.
The spatial dimension is generalized to  $d = 3-2\epsilon$ dimensions.
Integrals are evaluated at a value of $d$ for which they converge and then
analytically continued to $d=3$.
We use the integration measure
\begin{equation}
 \int_p\;\equiv\;
  \left(\frac{e^\gamma\mu^2}{4\pi}\right)^\epsilon\;
\:\int {d^{3-2\epsilon}p \over (2 \pi)^{3-2\epsilon}}\;.
\label{int-def}
\end{equation}

We require a few integrals in that appear in the soft sector.
The momentum scale in these integrals is set by the Debye mass $m$.
The one-loop integrals needed are:
\bqa
\int_p\log\left(p^2+m^2\right)&=&
-{m^3\over6\pi}
\;,
\label{1l0} 
\\ 
\int_p{1\over p^2+m^2}&=&
-{m\over4\pi}
\;.
\label{1l1}
\eqa

The errors are all of order $\epsilon$.
The calculation of these integrals is standard.


\end{document}